\documentclass[letterpaper,fleqn,usenatbib]{mnras}

\usepackage{enumerate}

\def\cf{{\it cf.}}
\def\eg{{\it e.g.}}
\def\etal{{\it et al.}}

\def\ie{{\it i.e.}}

\def\Msun{M$_\odot$}

\def\pmb#1{\setbox0=\hbox{$#1$}%
  \kern-0.25em\copy0\kern-\wd0
  \kern.05em\copy0\kern-\wd0
  \kern-0.025em\raise.0433em\box0}
\def\spmb#1{\setbox1=\hbox{${\scriptstyle #1}$}%
  \kern-0.25em\copy1\kern-\wd1
  \kern.05em\copy1\kern-\wd1
  \kern-0.025em\raise.0433em\box1}

\long\def\Ignore#1{\relax}
\usepackage{color}
\definecolor{red}{rgb}{0.7,0.1,0.1}
\definecolor{blue}{rgb}{0.2,0.2,0.8}
\definecolor{green}{rgb}{0.1,0.6,0.1}

\usepackage{graphicx}	% Including figure files
\title[Spiral instabilities]{Spiral instabilities: Linear and nonlinear effects}

\author[Sellwood \& Carlberg]
          {J. A. Sellwood,$^{1}$\thanks{E-mail:sellwood@as.arizona.edu}
and
{R. G. Carlberg,$^2$\thanks{E-mail: raymond.carlberg@utoronto.ca}}
\\
$^1$Steward Observatory, University of Arizona, 933 N Cherry Ave,
Tucson AZ 85722, USA \\ $^2$Department of Astronomy and Astrophysics,
University of Toronto, ON M5S 3H4, Canada}

\pubyear{2020}

\begin{document}
\label{firstpage}
\pagerange{\pageref{firstpage}--\pageref{lastpage}}
\maketitle

\begin{abstract}
We present a study of the spiral responses in a stable disc galaxy
model to co-orbiting perturbing masses that are evenly spaced around
rings.  The amplitudes of the responses, or wakes, are proportional to
the masses of the perturbations, and we find that the response to a
low-mass ring disperses when it is removed -- behaviour that is
predicted by linear theory.  Higher mass rings cause nonlinear changes
through scattering at the major resonances, provoking instabilities
that were absent before the scattering took place.  The separate wake
patterns from two rings orbiting at differing frequencies, produce a
net response that is an apparently shearing spiral.  When the rings
have low mass, the evolution of the simulation is both qualitatively
and quantitatively reproduced by linear superposition of the two
separate responses.  We argue that apparently shearing transient
spirals in simulations result from the superposition of two or more
steadily rotating patterns, each of which is best accounted for as a
normal mode of the non-smooth disc.

% 165 words
\end{abstract}


%%%%%%%%%%%%%%%%% BODY OF PAPER %%%%%%%%%%%%%%%%%%

\section{Introduction}
\label{sec.intro}
In earlier papers \citep{SC14, SC19}, we argued that spiral patterns
in galaxies can result from a recurrent cycle of instabilities in the
stellar disc.  Each instability is excited by a deficiency in the
distribution of stars over a narrow range of angular
momentum, that was created by resonant scattering by a previous
disturbance.  The cycle can be initiated by the infall of a mass clump
during disc assembly, say, or by a tidal interaction, or in the
unlikely event that neither of these triggers occurs, it can bootstrap
out of the noise \citep{Se12}.  Resonant scattering by successive
waves raises the level of random motion wherever it occurs, making the
surviving disc gradually less responsive so that the recurrrent cycle
is self-limiting in the absence of a dissipative gas component
\citep{SC84, CF85, To90, Ro08, Au16}.

The vigorous instabilities we invoke are normal modes of the modified
disc.  They can be calculated by first order perturbation theory
\citep{Ka71, SK91} that assumes an infinitesimal perturbation and
neglects possible changes to the background state.  A linear
instability will grow from any arbitrarily small seed perturbation,
but the exponential growth must saturate at some finite amplitude due
to the breakdown of the assumption that second- and higher-order
terms can be neglected.  Scattering at resonances is another
nonlinear effect that changes the background state of the disc by
wave-particle interactions that are strictly second order \citep{LBK,
  Ma74}.  Thus our proposed mechanism invokes both linear perturbation
theory to account for the growth of the spirals, and nonlinear
scattering at resonances to account for the seeding of subsequent
instabilities.

Swing-amplification \citep{GLB, To81, BT08}, which also causes wakes
to form around co-orbiting perturbing masses \citep{JT66, Bi20}, is
another important result from linear perturbation theory in stellar
discs in which second order changes are also neglected.  But it is not
an instability, in the sense of a normal mode, and it merely amplifies
an input perturbation by a finite, sometimes large, factor.  However,
scattering of stars caused by the ultimate resonant damping of the
disturbance \citep{To81} may change the stability properties of the
entire disc \citep{Se12}, even for tiny perturbations.

A number of authors \citep{Wada11, Gran12a, Gran12b, Roca13, Ka14,
  DB14, Baba15, MK18, MK20} argue for an alternative picture, in which
spiral patterns are hardly density waves at all, but wind more tightly
over time at a rate that is almost as rapid as if they were material
features.  These authors find evidence from their simulations that
swing-amplification plays a prominent role in the development of the
spirals and some aspects of the behaviour are attributed to
nonlinearities \citep{Baba13, KN16}.  

A third mechanism for the origin of spirals is argued by \citet{To90},
\citet{TK91}, and \citet{DVH}.  These authors propose that galaxies do
not possess spiral instabilities at all, and that the patterns are no
more than the collective responses of the surrounding disc to
co-orbiting density inhomogeneities, such as giant molecular clouds
perhaps.  The linear theory of this behaviour was originally set out
by \citet{JT66}, but \citet{DVH} show that massive enough clumps in a
low-mass disc can trigger nonlinearities in the response that lead to
continuing development of patterns after the seed mass has been
removed.

Thus all three of these proposed mechanisms for spiral generation
invoke nonlinearities.  Here we endeavour to clarify the distinction
between linear and nonlinear behaviour in disc galaxy simulations.  We
present experiments that employ a stable disc model and have large
enough particle numbers that the evolution of mild perturbations
reflects the predictions of linear theory, while larger disturbances
behave differently.  

Our purpose in presenting these simplified experiments is to shed
additional light on the behaviour of self-gravitating disturbances in
shearing discs.  Our idealized experiments are more easily understood
than those bearing a closer resemblance to real galaxies and
illustrate clearly how visual impressions can be misleading.

We do not dispute the numerical results of others, but present an
alternative interpretation of shearing spirals in \S\ref{sec.shear}.
We also confirm one result from \cite{DVH} in \S\ref{sec.DVH0}, but
show that wakes in more massive discs sculpt the distribution function
in a manner to excite new instabilities.  Thus we find that the only
nonlinearity one needs to invoke to explain the appearance of fresh
disturbances seeded by a larger perturbation is scattering of
particles at the major resonances of the original disturbance.

\section{Technique}
\label{sec.methods}
As we noted before \citep{SC19}, the dynamical behaviour of fully
self-consistent simulations can be hard to unravel, and we therefore
find it fruitful to run simplified simulations that can capture the
phenomena we wish to study without them being obscured by unrelated
activity.  We therefore first choose a model that has been proven to
be globally stable and study its response to controlled perturbations,
while restricting disturbance forces to a single sectoral harmonic.
We present a different model in \S\ref{sec.DVH0}.

\subsection{Mestel disc model}
Once again, we adopt the razor-thin Mestel disc used in the
studies by \citet{Za76} and \citet{To81}, which is characterized by a
constant circular speed $V_0$ at all radii.  The surface density of
the full-mass disc is $\Sigma_0(R) = V_0^2/(2\pi GR)$.

\begin{figure}
\includegraphics[width=.9\hsize,angle=0]{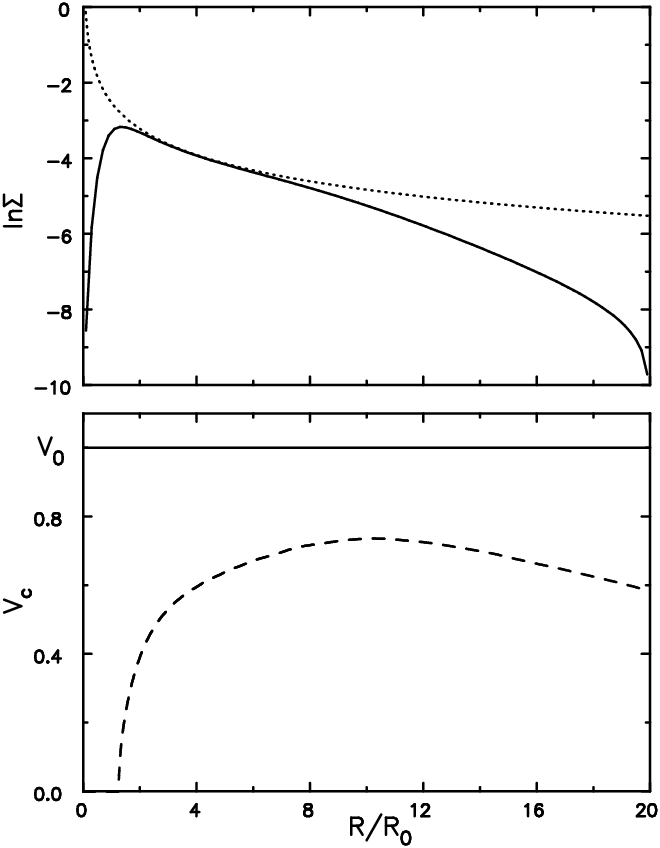}
% /home/sellwood/papers/SC20/figs/Mestel/dflook.s
\caption{Above: the solid curve gives radial variation of the surface
  density of the particles in the tapered half-mass Mestel disc.  The
  dotted curve shows what the surface density would be without the
  tapers.  Below: the horizontal solid line shows the (constant)
  circular speed in the Mestel disc, the dashed curve is the
  contribution from the disc particles.}
\label{fig.Mestel}
\end{figure}

The advantage of using this model is that \citet{To81} reported that
the half-mass Mestel disc, with $Q=1.5$ and a suitable central cut out
had no linear instabilities whatsoever.  Thus an $N$-body realization
having a large number of particles to beat down shot noise, should not
evolve.  Tests of this prediction \citep{Se12} confirmed that it was
true for quite a long time, though he also reported that instabilities
eventually appeared in very long integrations.  Here we do not run our
simulations for nearly long enough for this subtle change of behaviour
to affect the results we report.

The particle distribution function (DF), tapers and cut-offs are the
same as in eqs.~(4) -- (6) of \citet{SC19} and we again adopt units
such that $V_0 = R_0 = G = 1$, where $R_0$ is the central radius of
the inner cut out.  As before we halve the disc mass, and maintain
centrifugal equilibrium by keeping the central attraction $a_R =
-V_0^2/R$ at all times.  Figure~\ref{fig.Mestel} gives the surface
density of the active mass (solid line, upper panel); the tapers
change the power law profile (dotted curve) to one that is
approximately exponential with a central hole.  The solid horizonal
line in the lower panel is the circular speed; the contribution that
the particles would make is drawn by the dashed curve, indicating
that our disc is significantly sub-maximal.

The natural unit of mass of the full-mass Mestel disc is $V_0^2R_0/G$,
which is the mass enclosed within $R_0$, and the disc mass would
increase linearly with $R$ in the absence of tapers.  The total mass
of the particles in our model, after allowing for the various tapers,
truncations and halving of the active mass, is $M_{\rm act}\simeq5.40$.

\begin{table}
\caption{Numerical parameters}
\label{tab.params}
\begin{tabular}{@{}ll}
Grid size & 106 $\times$ 128 \\
Active sectoral harmonic & 3 \\
$R_0$ & 8 grid units \\
Softening length & $R_0/8$ \\
Number of particles & $5\times 10^7$ \\
Basic time-step & $R_0/(80V_0)$ \\
Time step zones & 5 \\
Guard zones & 4 \\
\end{tabular}
\end{table}

\subsection{Numerical method}
The particles in our simulations are constrained to move in a plane
over a 2D polar mesh at which the self-gravitational attractions are
calculated and interpolated to the position of each particle.  A full
description of our numerical procedures is given in the on-line manual
\citep{Se14} and the code itself is available for download.
Table~\ref{tab.params} gives the values of the numerical parameters
for all the simulations presented in
\S\S\ref{sec.no-mass}--\ref{sec.tworings}.  \citet{Se12} reported, and
we have reconfirmed in this study, that all our results are
insensitive to reasonable changes to grid resloution, time step and
zones, and number of particles.  Doubling the softening length does
reduce the responsiveness of the disc, but does not alter the
qualitative behaviour.

Since the gravitational field is a convolution of the mass density
with a Green function that is most efficiently computed by Fourier
transforms, it is easy to restrict the sectoral harmonics that
contribute to the field when using a polar grid.  In the simulations
we report here, the disturbance forces arising from the particles are
confined to just the $m=3$ sectoral harmonic.  Imposing 3-fold
symmetry implies slightly more vigorous swing amplification
\citep[][their Figure 6.21]{To81, BT08} than for $m=2$ because $X=4/m$
in this half-mass Mestel disc.

Not only does our choice of a grid code enable us readily to select
active sectoral harmonics, but it is very much faster than tree codes.
\citet{Se14} reported performance benchmarks and also demonstrated that
the computed forces acting on each particle are very nearly the same
in the two codes.  Our method is admittedly less versatile than tree
codes, but is ideally suited for the evolution of isolated
collisionless models.  For this application, its perfomance advantage
over tree codes can be likened to that of FFTs over the direct FTs --
it gets the same answer in a fraction of the cpu time.  Furthermore,
our code has been validated by verification of the normal modes
predicted by lineary theory \citep{SA86}, and for non-linear evolution
\citep{INS84}, and confirms the predicted stability of the half-mass
Mestel disc \citep[][and this paper]{To81, Se12}.

\subsection{Other details}
The models we present in \S\ref{sec.results} generally include three
co-orbiting masses equally spaced around a ring of radius $R_p$, and
in some cases we employ two such rings at differing radii.  The
perturbing masses were each Plummer spheres \citep{BT08} with masses
$M_p$, a parameter we vary, and core radius $a=0.05R_0$.  The
attraction of these perturbing masses on each disc particle is
computed directly, and added to the non-axisymmetric forces from the
other particles, computed through the grid, plus the fixed central
attaction.  The ring particles are moved at each step at fixed speed
$V_0$ around a circle of radius $R_p$.

Table \ref{tab.sims} lists all the simulations reported in
\S\S\ref{sec.no-mass}--\ref{sec.tworings}.

We will use the radial action of a particle, $J_R \equiv \oint \dot R
dR/(2\pi)$ \citep{BT08}, which has dimensions of angular momentum.  It
is zero for a circular orbit, and increases with the eccentricity of
the orbit.  In an axisymmetric potential, the angular momentum, $L_z$,
is the other action for orbits confined to a plane.

We also make extensive use of logarithmic spiral transforms of the
particle distribution, which for $N$ particles of equal mass, as here,
is defined as
\begin{equation}
A(m,\tan\gamma,t) = {1 \over N}\sum_{j=1}^N \exp[im(\phi_j + \tan\gamma \ln R_j)],
\label{eq.logspi}
\end{equation}
where $(R_j,\phi_j)$ are the cylindrical polar coordinates of the
$j$-th particle at time $t$, and $\gamma$, the complement to the
spiral pitch angle, is the angle between the radius vector and the
tangent to an $m$-arm logarithmic spiral, with positive values for
trailing spirals.  We usually plot the time evolution of the amplitude
$|A|$, which ignores the complex phase.  If the particles were
randomly distributed, the expectation value of $\langle A^2 \rangle =
1/N$, while $\langle|A|\rangle = [\pi/(4N)]^{1/2}$, for any $m$ and
$\tan\gamma$ \citep{Ma72}.

\begin{figure}
\includegraphics[width=.9\hsize,angle=0]{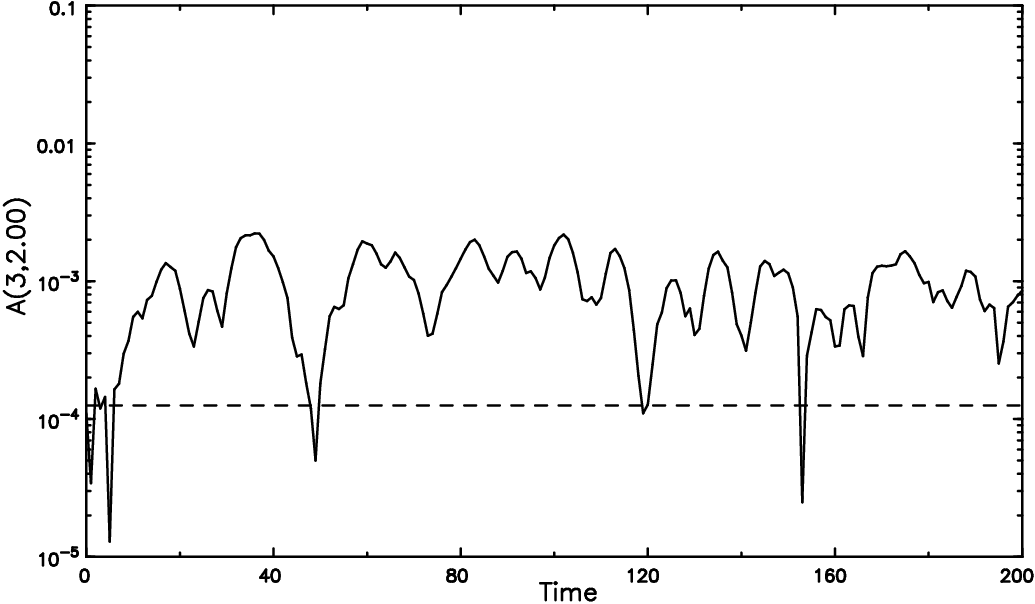}
% /home/sellwood/blobs/no-mass.s
\caption{The time evolution of the $|A(3,2)|$ component of the
  transform (eq.~\ref{eq.logspi}) in model 0, a simulation with no
  forcing masses.  The large, short-term variations are explained in
  the text.  The dashed line is the expectation value for $N$ randomly
  distributed particles.}
\label{fig.no-mass}
\end{figure}

\begin{table*}
\caption{List of simulations with three-fold perturbations in the Mestel disc}
\label{tab.sims}
\begin{tabular}{@{}ll}
Model 0 & No forcing masses \\
Model 1L & One ring at $R_p=3$ with $M_p = 2 \times 10^{-4}$  \\
Model 1LR & Model 1L continued from $t=50$ with the driving ring removed \\
Model 1H & One ring at $R_p=3$ with $M_p = 10^{-3}$  \\
Model 1HR & Model 1H continued from $t=50$ with the driving ring removed \\
Model 1HRS & As for Model 1HR, but with the particle coordinates scrambled at $t=50$ \\
Model 1HSp & Model 1HR, but with DF spliced to remove scattering features from the Lindblad resonances \\
Model 2L & Two rings at $R_{p,1}=2.5$ and $R_{p,2}=4$ with $M_p = 10^{-4}$ \\
Model 2H & Two rings at $R_{p,1}=2.5$ and $R_{p,2}=4$ with $M_p = 10^{-3}$ \\
\end{tabular}
\end{table*}

\section{A simulation without forcing}
\label{sec.no-mass}
Although a smooth stellar fluid model of this disc is globally stable
\citep{To81}, the shot noise in any particle realization induces a
swing-amplified response.  We therefore first report model 0 to
calibrate the evolution of this 50M particle disc in the absence of
any driving masses when, for the reasons given in \S\ref{sec.methods},
disturbance forces were restricted to $m=3$ only.

Figure~\ref{fig.no-mass} shows the time evolution of the amplitude of
the $|A(3,2)|$ component of the transform (\ref{eq.logspi}).  This
trailing spiral component has a pitch angle $(90 - \gamma) \simeq
27^\circ$, which is close to the peak of the swing-amplified response,
but the behaviour of other trailing components in the range $1 \leq
\tan\gamma \leq 3$ is very similar.  While manifesting large
variations with time, $|A(3,2)|$ is generally many times higher than
that expected from a random distribution, shown by the dashed line.
This is because the self-gravity of the disc causes swing-amplified
collective responses to the density variations seeded by shot noise;
in effect, self-gravity polarizes the particle distribution, as
described in detail by \citet{TK91}.  The fluctuations in the net
spiral amplitude that are evident in Figure~\ref{fig.no-mass} result
from destructive and constructive interference between different
noise-driven spiral segments that orbit at differing angular
frequencies at different radii.  The mean excess over the random
expectation value depends on the responsiveness of the disc, which is
determined by the swing-amplification parameters $X$, $Q$, the
dimensionless shear rate $\Gamma \equiv -d\ln \Omega_c/d\ln R$, as
well as the softening length.

\begin{figure*}
\includegraphics[width=.6\hsize,angle=270]{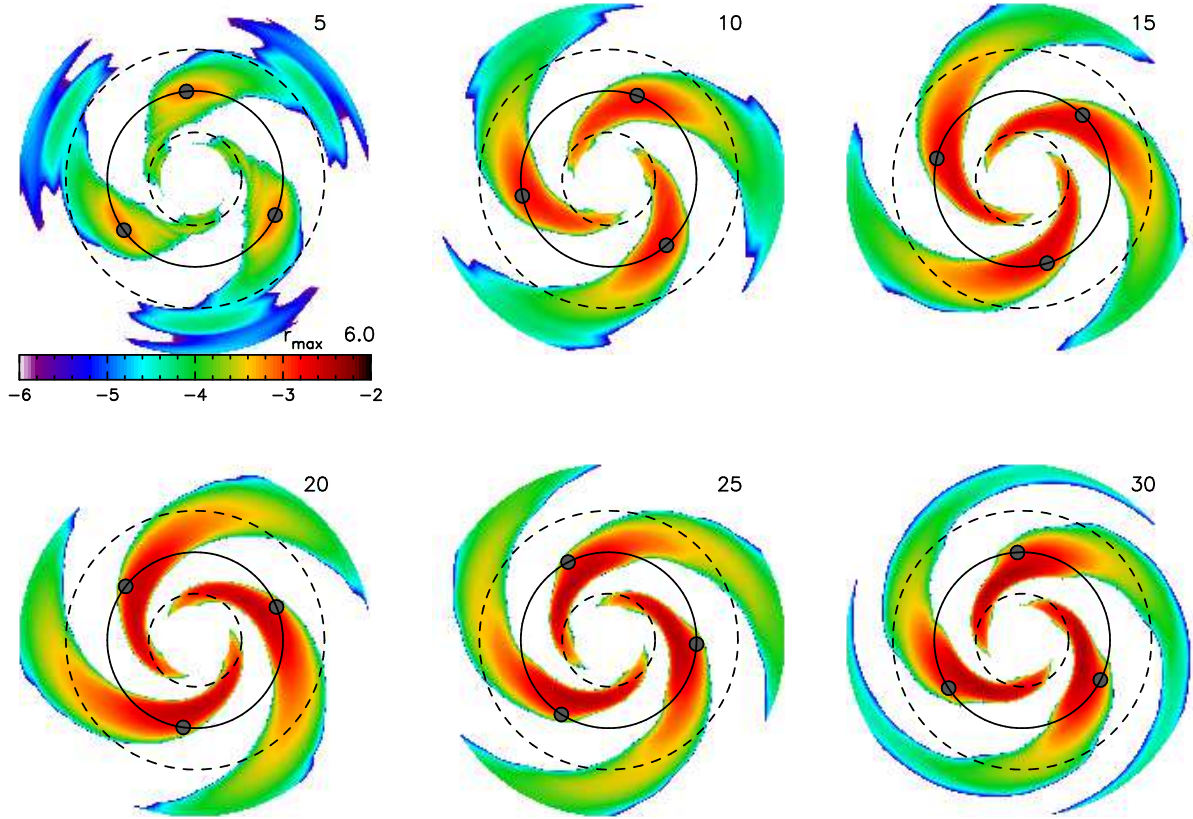}
% /home/sellwood/blobs/show-den.s 4948
\caption{The growth of the relative response density to a single ring
  of perturbing masses, marked by the large grey dots.  Only the
  positive parts of the response density, on the given logarithmic
  scale, and over the range $1.2<R<6$ are shown, and the figure is
  drawn for model 1H, the case of the heavier ring because it is less
  contaminated by noise.  The full-drawn circle marks the corotation
  resonance and the dashed circles mark the Lindblad resonances.}
\label{fig.show-den}
\end{figure*}

Because of the large number of particles, the amplified responses are
still quite mild, and remain in the linear regime for a long time,
during which any secular growth is slow.  We do not show the evolution
of model 0 beyond $t=200$, but have continued it to $t=1000$, finding
that the amplitude rises rapidly later due to nonlinear resonant
scattering and three-arm spiral patterns become visible.  This
behaviour is similar to that reported by \citet{Se12} for $m=2$ forces
in the same model with the same file of initial particles.  None of
the forced simulations we report here is affected by this late time
nonlinear change in the behaviour because they were run to $t=200$
only.

\begin{figure}
\includegraphics[width=.9\hsize,angle=0]{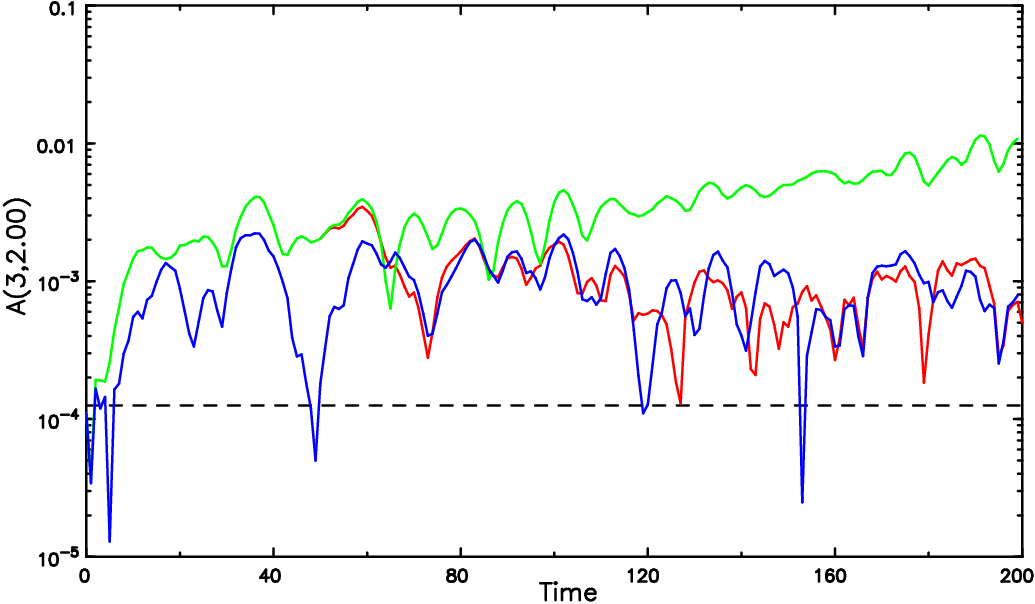}
% /home/sellwood/blobs/lo-mass.s
\caption{The time evolution of the $|A(3,2)|$ component of the
  transform (eq.~\ref{eq.logspi}) in three simulations. The green
  line, from model 1L, is when the low-mass perturbing ring at $R_p=3$
  is retained throughout, the red (model 1LR) is when the ring is
  removed at $t=50$, and the blue line (model 0) and dashed line are
  reproduced from Fig.~\ref{fig.no-mass}.}
\label{fig.lo-mass}
\end{figure}

\section{Experiments with a single ring of perturbing masses}
\label{sec.results}
A few snapshots from the evolution of model 1H are shown in
Figure~\ref{fig.show-den}, which report the early growth of the
density response to an imposed ring of heavy co-orbiting particles,
marked by the three dots in each frame.  As already noted, disturbance
forces were restricted to the $m=3$ sectoral harmonic.  The orbit
period of the individual masses is $6\pi$ in our units.  This Figure
is drawn for model 1H, used in \S\ref{sec.results}.2 below because the
response to this heavier ring stands out more clearly from the noise,
although even in this case the peak fractional over-density of the
response is less than 1\% by $t=30$.

\subsection{A low-mass ring}
Figure~\ref{fig.lo-mass} presents the time evolution of a trailing
component amplitude from several simulations.  The blue curve is from
model 0, and is reproduced from Figure~\ref{fig.no-mass}.  The green
line reports the amplitude evolution in model 1L with a single ring of
three perturbing masses at $R_p=3$, each having a mass $M_p = 2 \times
10^{-4}$, which is $3.7 \times 10^{-5}$ of the total disc mass or that
of $\sim 1\,852$ particles.  The response to this trefoil perturbation
rises quickly at first, just as in the unperturbed disc (blue line),
because the physics of the response to this type of forcing is
identical to that of disc polarization but, because the
perturbation is substantially heavier than a single particle, the
overall amplitude is a little higher, making the interference from the
noise-driven fluctuations relatively weaker.  Consistent with
Figure~\ref{fig.show-den}, the growth of the response slows after a
couple of orbits of the perturbing ring.  Linear theory \citep{JT66}
predicts that the response to a single perturbing mass asymptotes to a
steady value after about 5 epicyclic periods, which would be by
$t\sim65$ in this case, although it is clear that the amplitude in our
model continues to rise slowly to the last moment shown.

The ring of three perturbing masses was present throughout the
simulation shown by the green line.  In a separate experiment, model
1LR, we removed the perturbing ring of particles at $t=50$, or after
2.65 orbits, and the amplitude evolution from this time is shown by
the red line.  As linear theory would predict, the response amplitude
begins to decrease as soon as the perturbation is removed, and the red
line quickly decays back to the level in model 0, indicated by the
blue line.

\subsection{A heavier ring}
\label{sec.1H}
\begin{figure}
\includegraphics[width=.9\hsize,angle=0]{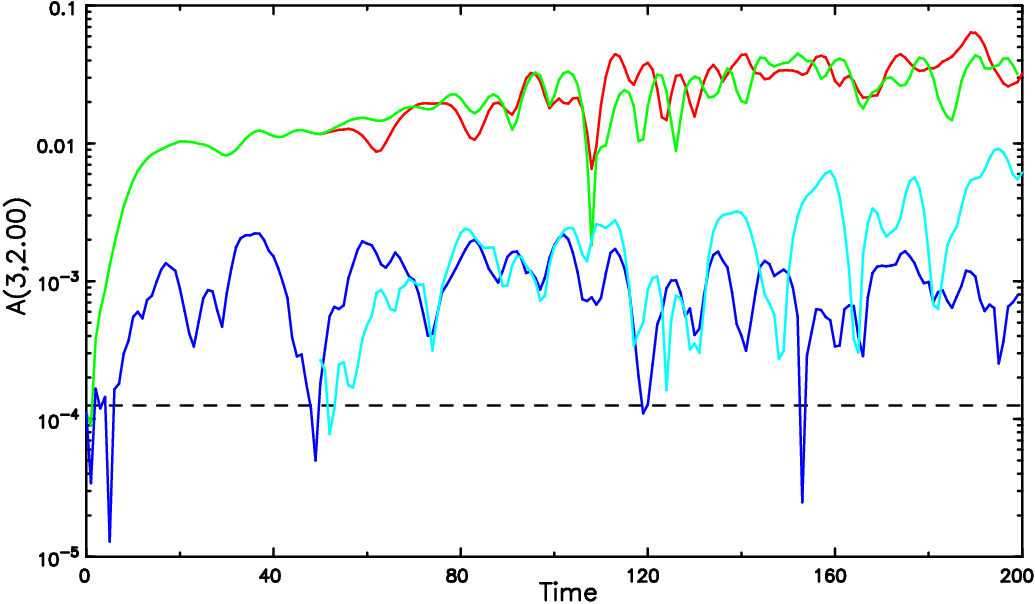}
% /home/sellwood/blobs/hi-mass.s
\caption{The time evolution of the $|A(3,2)|$ component of the
  transform (eq.~\ref{eq.logspi}). The green line is from model 1H in
  which the high-mass perturbing ring at $R_p=3$ was retained
  throughout, the red is from model 1HR in which the ring was removed
  at $t=50$, the cyan line is from model 1HRS in which the perturbing
  ring has been removed and the particle coordinates scrambled in
  azimuth to erase the existing density response, while the blue and
  dashed lines are reproduced from Fig.~\ref{fig.no-mass}.}
\label{fig.hi-mass}
\end{figure}

The amplitude evolution in model 1H, a run with five times heavier
perturbing masses, is shown in Figure~\ref{fig.hi-mass}.  At early
times, the response amplitude is close to five times that in model 1L
(Figure~\ref{fig.lo-mass}), consistent with linear theory expectations.
However, the amplitude given by the red line, model 1HR, indicates
that the overall response did not decay in this separate experiment in
which the heavier perturbing ring was removed, but continued to rise
pretty much as in model 1H when the perturbing masses remained
present (green line).

\begin{figure}
\includegraphics[width=.9\hsize,angle=0]{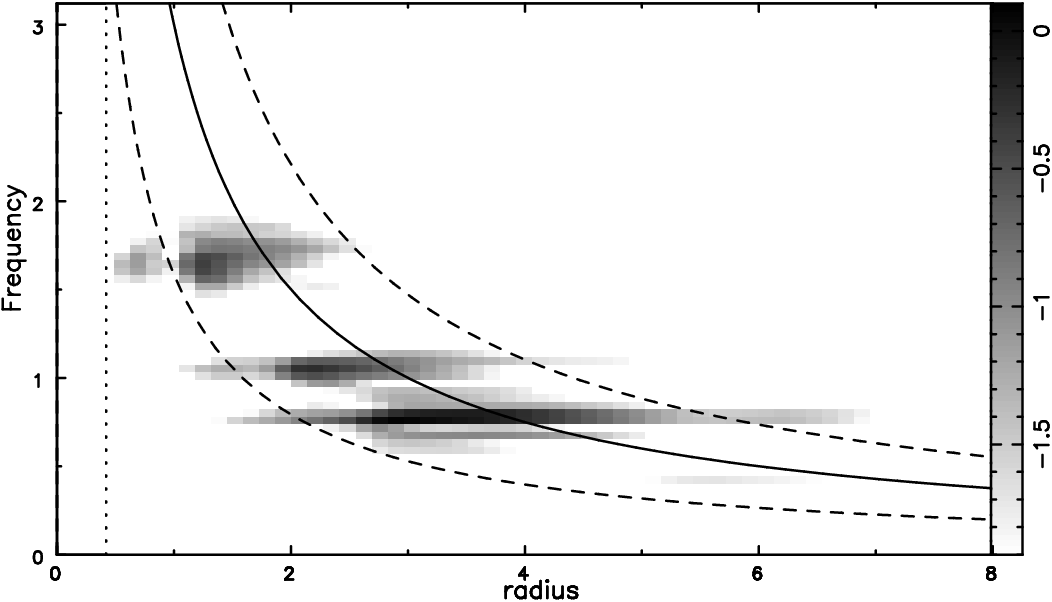}
% /home/sellwood/blobs/hi-spct.s
\caption{The grey scale shows the logarithm of power in the $m=3$
  sectoral harmonic of the disturbance density over the time interval
  $51\leq t\leq 200$ in model 1HR after the heavier forcing ring was
  removed. The solid curve indicates the circular angular frequency
  $m\Omega_c$, and the dashed curves $m\Omega_c\pm\kappa$.  Data
  interior to the dotted line were excluded.  Each horizontal streak
  indicates a disturbance of fixed pattern speed.}
\label{fig.hi-spct}
\end{figure}

\begin{figure}
\includegraphics[width=.9\hsize,angle=0]{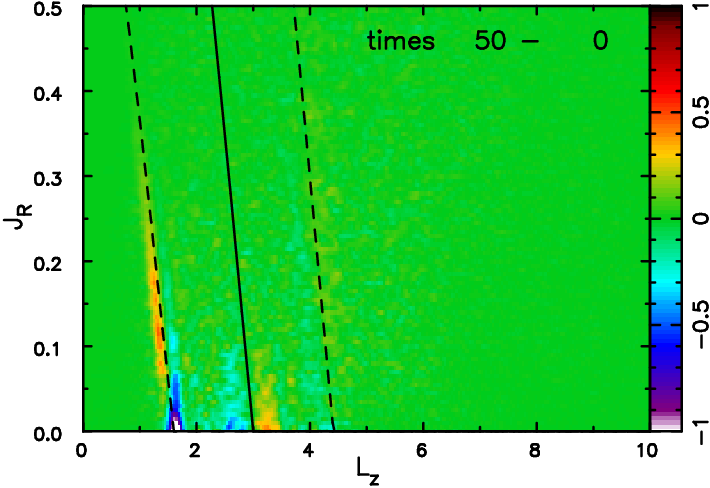}
% /home/sellwood/work/hi-diff.s
\caption{The difference between density of particles in action space
  at $t=50$ and $t=0$ in model 1H.  The lines record the loci of the
  major resonances of the driving ring: full-drawn for corotation and
  dashed lines for the Lindblad resonances.}
\label{fig.hi-diff}
\end{figure}

The response amplitude to this heavier ring rises at first to an
almost steady value, because it stands out more clearly from the
noise, but fluctuations in the green (model 1H) and red (model 1HR)
lines in this figure again develop after $t\sim50$, which reflect
beats between multiple disturbances of comparable amplitude having
differing angular frequencies.  The power spectrum from the period
after the ring was removed in model 1HR, displayed in
Figure~\ref{fig.hi-spct}, reveals three disturbances having distinct
pattern speeds.  That with $m\Omega_p \approx 1$ has the frequency of
the original forcing ring with corotation at $R=3$, and Lindblad
resonances at $R\approx1.6$ and $R\approx4.4$.  The other two
frequencies must have been excited by other means and, since
corotation for each of these disturbances is close to the Lindblad
resonance of the original wave, it seems likely they are groove modes
excited by deficiencies in the DF created by Lindblad resonance
scattering.  This conjecture is confirmed in Figure~\ref{fig.hi-diff},
which shows changes by $t=50$ in model 1H to the density of particles
in the space of the two actions, $L_z$ and $J_R$.  Forcing by the
heavier ring has scattered particles to larger $J_R$ at the Lindblad
resonances, marked by the dashed lines.  The deficiencies of particles
created at low $J_R$ near $L_z=1.6$ and, less distinctly, at
$L_z\approx4.3$, each excite a new instability.

Figure~\ref{fig.hi-diff} also reveals that some particles have crossed
corotation, which is reminiscent of radial migration \citep{SB02}.
The deficiency for small $J_R$ at $L_z<3$ and excess at $L_z>3$
indicates there were greater numbers of outward movers over inward,
which is consistent with the negative gradient $\partial f/\partial
L_z|_{J_R}$ near $L_z=3$ \citep[see][Fig.~1]{Se12}.  However, the
parallel with the process discussed by \citet{SB02} is not exact
because here the disturbance was the forced response to a co-orbiting
ring of perturbers.  The disc response to constant forcing grows
quickly at first (green line in Figure~\ref{fig.hi-mass}), but becomes
almost steady well before $t=50$; changes to home radii at one end of
a horseshoe orbit are exactly undone at the other end when the
disturbance amplitude remains constant; in effect the particle is
trapped in the resonance.  Thus the net changes near corotation can
have occurred only as the particles became trapped.

We have used our mode fitting software \citep{SA86} to estimate the
frequencies of the three disturbances in model 1HR visible in
Figure~\ref{fig.hi-spct}, finding $\omega_1 \approx 1.054 - 0.012i$,
$\omega_2 \approx 1.66 + 0.001i$, and $\omega_3 \approx 0.772 +
0.015i$.  The real parts of these frequencies are the angular rotation
rates, which are in good agreement with the frequencies indicated in
Figure~\ref{fig.hi-spct}.  However, the imaginary parts, which
correspond to the growth rates, are of doubtful significance; that for
disturbance 2 is probably an underestimate, because the instability
had almost saturated over the time interval of the fit.  Note also
that although we have fitted a ``mode'' to disturbance 1, it should
not be thought of as a mode, since it is the decaying response to a
driving perturbation that was suddenly removed.  An imaginary part to
$\omega_1$ was required by the functional form adopted for the fit,
and while the negative value indicates decreasing amplitude, in
reality the decay is not exponential.

\begin{figure}
\includegraphics[width=.9\hsize,angle=0]{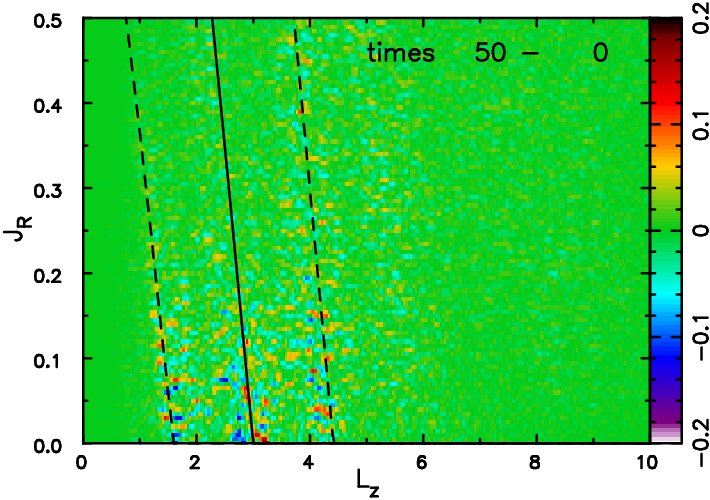}
% /home/sellwood/work/lo-diff.s
\caption{The difference between the density of particles in action
  space at $t=50$ and $t=0$ in model 1L having the lighter ring
  perturbation.  Note the difference in the color scale from that in
  Fig.~\ref{fig.hi-diff}}
\label{fig.lo-diff}
\end{figure}

Figures \ref{fig.hi-diff} and \ref{fig.lo-diff} compare the changes in
action space density in model 1H, the heavier ring, with that in model
1L having the lighter ring.  It is important to notice the five-fold
difference in the colour scale between the two figures.  The lighter
ring indeed causes some very mild resonance scattering, but it is
evidently too weak to excite new modes, and the density response
simply decays away (Fig.~\ref{fig.lo-mass}).  This is a clear
illustration of the difference between linear and nonlinear
behaviour.

\begin{figure}
\includegraphics[width=.9\hsize,angle=0]{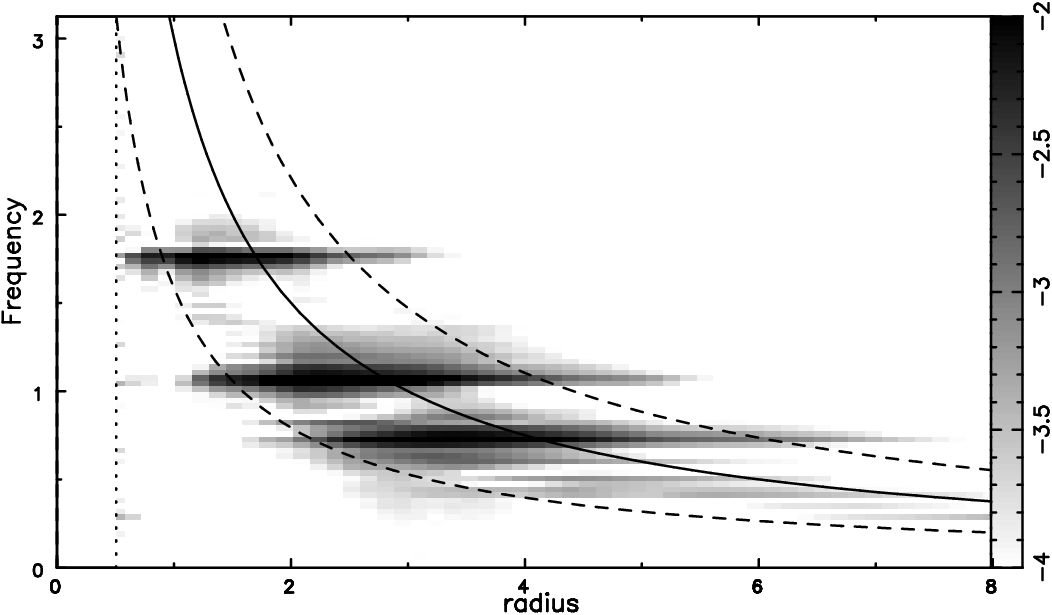}
% /home/sellwood/blobs/hi-scrb.s
\caption{As for fig.~\ref{fig.hi-spct}, but for model 1HRS in which
  the particle azimuths were also scrambled at $t=50$.}
\label{fig.hi-scrb}
\end{figure}

The cyan line in Figure \ref{fig.hi-mass} reports the amplitude
evolution in model 1HRS in which the forcing ring was not only
removed, but the particle azimuths were rearranged at random in order
to disperse the wake response to the imposed ring.  The power spectrum
from model 1HRS, shown in Figure~\ref{fig.hi-scrb}, reveals at
least three coherent waves; not just the two groove modes excited at
the Lindblad resonances of the original forced ring but also,
unexpectedly, a disturbance with $m\Omega_p \ga 1$, which was close to
the original forcing frequency.  Figure~\ref{fig.hi-diff} manifested
changes in the phase space density across corotation, and it seemed
possible that the deficiency at slightly lower $L_z$ than the
resonance had also seeded an instability.

\begin{figure}
\includegraphics[width=.9\hsize,angle=0]{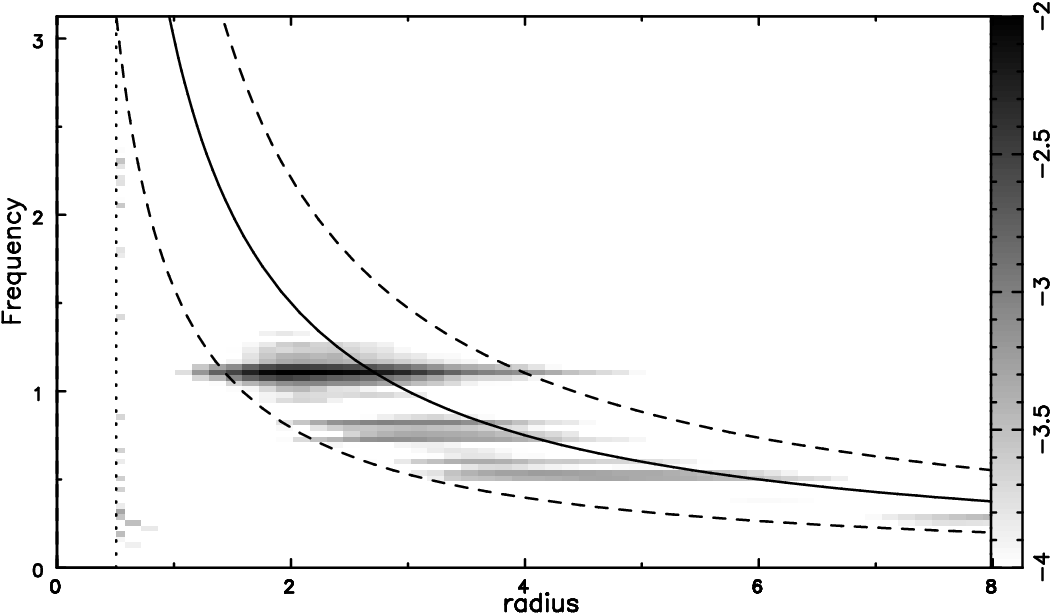}
% /home/sellwood/blobs/spl-spct.s
\caption{As for fig.~\ref{fig.hi-scrb}, but for model 1HSp in which the
  DF shown in fig.~\ref{fig.hi-diff} was spliced to retain only the
  features near corotation.  Now there is just one clear instability
  at very nearly the same frequency as the middle disturbance in
  fig.~\ref{fig.hi-scrb}.}
\label{fig.spl-spct}
\end{figure}

In order to check this possibility, we spliced the distribution of
particles near corotation into a pristine undisturbed DF, as we did in
\citet{SC19}, in order to suppress the instabilities associated with
the grooves created by Lindblad resonance scattering.  A simulation
with this modified DF, model 1HSp manifested just a single instability
standing out from the noise, as shown in Figure~\ref{fig.spl-spct}.
The fact that the pattern speed of this instability is very similar to
that of the middle disturbance in Figure~\ref{fig.hi-scrb}, and that
$m\Omega_p \ga 1$ in both cases confirms that the deficiency just
inside corotation in Figure~\ref{fig.hi-diff} also excites a groove
mode.

\subsection{Discussion}
The results reported in this section have clarified that a very
low-mass perturbation produces a linear response, that leads to no
lasting change.  The response decays when the driving masses are
removed (Fig.~\ref{fig.lo-mass}), and the changes to the DF caused by
the mild forcing by the low-mass ring (Fig.~\ref{fig.lo-diff}) were
not sufficiently strong or coherent to provoke new instabilities.  We
stress that this result could be obtained only if the disc has no
instabilities.  An externally imposed disturbance in a disc that
possesses even mild instabilities must always seed a growing response
that would not disperse were the driving perturbation to be removed.

The situation is quite different when the forcing masses were five
times heavier, though each was still only $0.017$\% of the disc mass.
The disc response was strong enough to cause coherent scattering at
the major resonances (Fig.~\ref{fig.hi-diff}), which in turn excited
instabilities where there had been none in the absence of the
perturbation.  The nature of the nonlinear response to this still
quite mild perturbation is now evident: scattering at the major
resonances created deficiencies in the DF at low-$J_R$, thereby
seeding new linear instabilities.

\begin{figure}
\includegraphics[width=.9\hsize,angle=0]{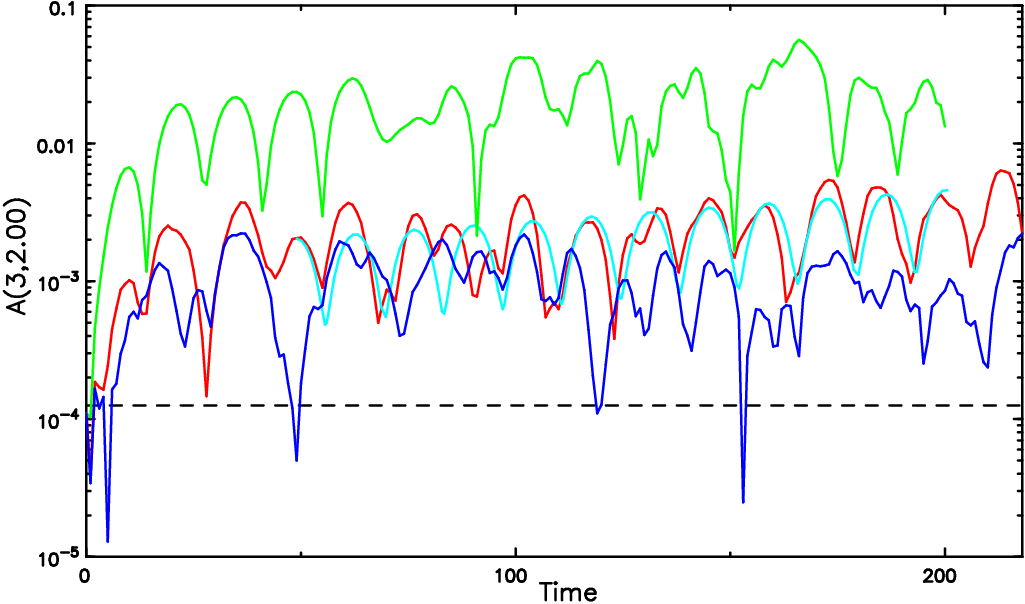}
% /home/sellwood/blobs/two-rings.s
\caption{As for fig.\ref{fig.lo-mass}, with the blue and dashed lines
  reproduced.  The red (model 2L) and green (model 2H) lines are for
  when two rings of forcing particles were applied; the red line is
  for the lower mass forcing particles while the green for ten times
  more massive particles.  The cyan line is described in the text.}
\label{fig.two-rings}
\end{figure}

\begin{figure}
\includegraphics[width=.9\hsize,angle=0]{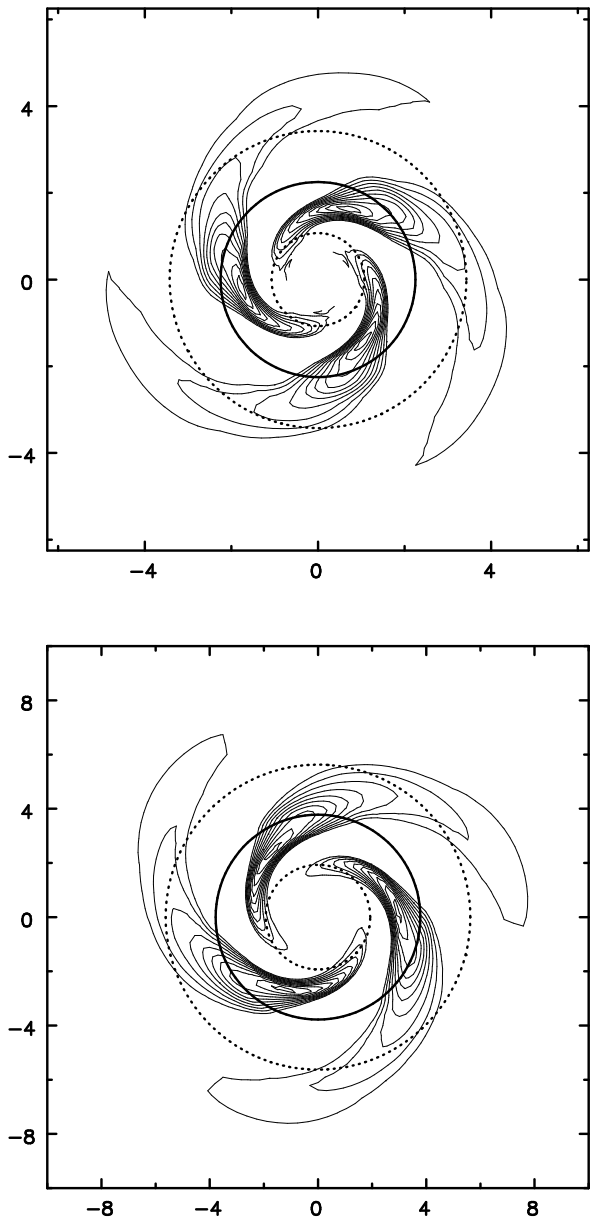}
% /home/sellwood/blobs/wavefit.s
\caption{The separate steady responses to the two low-mass rings of
  forcing particles fitted to the disturbance density in model 2L over
  the time range $50 \leq t \leq 200$.  The full-drawn circles mark
  corotation at $R\simeq 2.5$ (upper panel) and at $R\simeq 4$ (lower
  panel), while the dashed circles mark the Lindblad resonances for
  each disturbance.  Note that the spatial scale, in units of $R_0$,
  differs between the two panels in order that the corotation circle
  is the same size in each.}
\label{fig.waves}
\end{figure}

\begin{figure*}
\includegraphics[width=.65\hsize,angle=0]{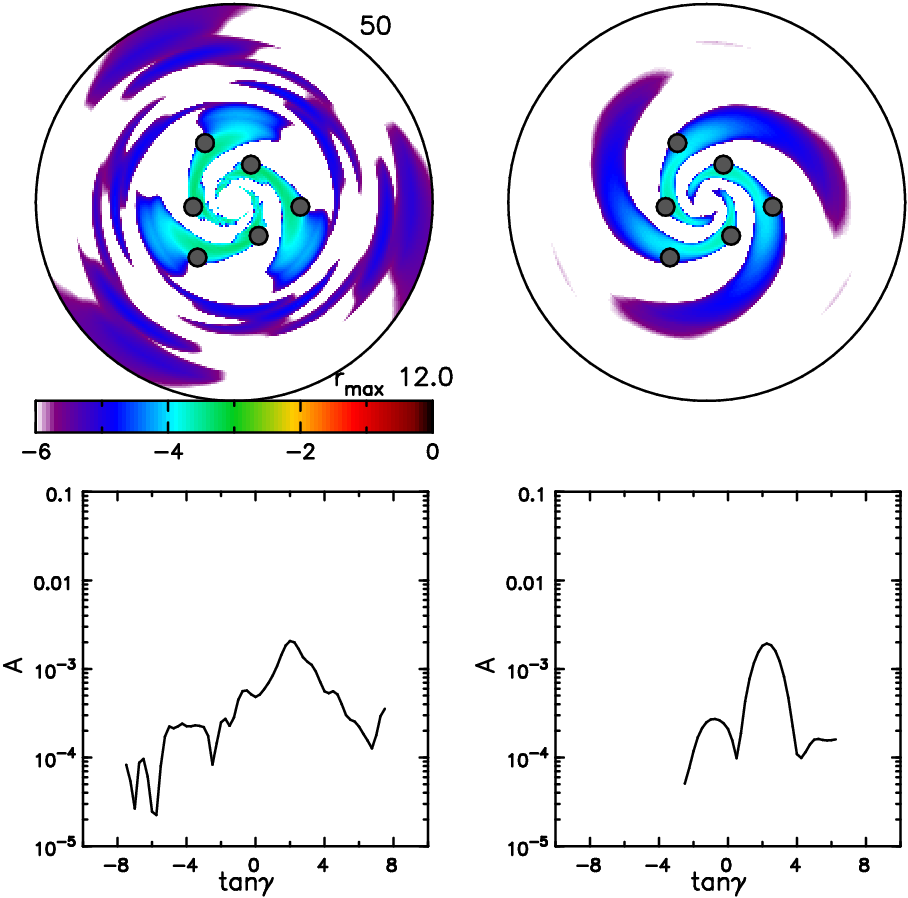}
% this place holder can be recreated with /home/sellwood/blobs/blobs.s 4945
%\includegraphics[width=.9\hsize,angle=0]{blobs4.avi}
\caption{An animation \textcolor{red}{(which will be included in the
    paper, but must be viewed separately for now)} showing the
  evolution of the disturbance density in the model 2L (top left) and
  its reconstruction from superposition of the two fitted waves from
  Fig.~\ref{fig.waves}.  The bottom row shows the evolution of the
  logarithmic spiral spectrum measured from the simulation (left) and
  its reconstruction from combining the fitted waves (right).  The
  comparison is very successful.}
\label{fig.movie}
\end{figure*}

\section{Experiments with two rings}
\label{sec.tworings}
We next report experiments in which we introduced two rings at radii
$R_{p,1}=2.5$ and $R_{p,2}= 4$.  In our first such case, model 2L,
each ring has three forcing particles of mass $M_p=0.0001$, \ie\ half
those in the lighter ring used in \S\ref{sec.results}.1, and we ran
the simulation to $t=200$ with this forcing continuously applied.  We
also present a second simulation, model 2H, having ten times heavier
forcing masses.

Figure~\ref{fig.two-rings} presents the time evolution of the
logarithmic spiral response in model 2L (red) and in model 2H (green).
Since the circular frequency at each ring is $\Omega_{c,1} = 0.4$ and
$\Omega_{c,2} = 0.25$, the beat frequency between the two disturbances
$m(\Omega_{c,1} - \Omega_{c,2}) = 0.45$ and the beat period is $\simeq
14.0$, which is consistent with the amplitude variations of the red
and cyan curves in the Figure, and of the green curve to $t\sim 50$.

% Mode  1 pattern speed    1.2120 growth rate    0.0054
% Mode  2 pattern speed    0.7553 growth rate    0.0056

\subsection{Lighter ring pair}
We have fitted two separate waves to the disturbance density in model
2L over the time interval $50<t<200$, finding $m\Omega_{p,1} = 1.212$
and $m\Omega_{p,2} = 0.755$, in good agreement with the two driving
frequencies.  The fit allowed for exponential growth of these waves
and returned the imaginary part of the frequency
$\Im(\omega)\simeq0.0055$ in both cases.  Since these disturbances are
forced responses and not modes, we do not attach much significance to
this growth rate that is $<1$\% of the real part of either frequency.

The positive parts of the fitted waves in model 2L are drawn in
Figure~\ref{fig.waves}, and in order emphasize that the two responses
are almost perfect scaled versions of each other, we changed the
spatial scale between the two panels so that the corotation circle
(full-drawn) has the same size in each.  This gratifying result stems
from the self-similarity of the Mestel disc, but can hold only where
the tapers to the infinite disc have little effect; recall that the
tapers were centered on $L_z=1$ and $L_z=11.5$, and will not much
affect the density of low-eccentricity orbits, that would dominate the
response, near either forcing ring.

The cyan line in Figure~\ref{fig.two-rings} is the superposition of
the logarithmic spiral components fitted to these two separate waves,
and its time evolution is in reasonable qualitative and quantitative
agreement with the red curve from model 2L.  We present a movie in
Figure~\ref{fig.movie} in which, in the top right panel, we
reconstruct the time evolution of the disturbance density as the
linear superposition of the two fitted waves, which is to be compared
with the density variations at each moment in the simulation in the
top left panel.  Notice in this movie how the superposed spiral arms
appear to wind up for a while before each breaks briefly into two
separate spiral segments that quickly rejoin to make a more open
pattern that winds again.  Each of the perturbing rings has a low
enough mass to ensure the response remains linear, as we showed above
for a single ring.  The close agreement in the movie between the time
evolution of model 2L and its reconstruction is consistent with linear
theory.

The lower two panels compare the logarithmic spiral spectrum measured
from the simulation, on the left, with the superposition of the fitted
transforms of each of the two separate waves on the right.  Here one
can see a repeating pattern of a peak moving to the right, from
leading to trailing as the amplitude rises to a maximum for an open
trailing spiral near $\tan\gamma=+2$, followed by a decrease as the
spiral continues to wind up.  This behaviour is reminiscent of
swing-amplification and the ``dust to ashes'' figure from
\citet{To81}, even though we know that the bottom right panel was
constructed by superposition of two steady waves.

A figure illustrating similar time evolution of the logarithmic spiral
transform in an unforced simulation was first presented by
\citet{SC84}, who interpreted it at the time as evidence that the
spiral pattern was caused by swing-amplification, as have many
subsequent authors.  However, \citet{Se89} later presented power
spectra of the same simulation, which showed that the apparently
shearing and swing-amplified pattern in fact resulted from the
superposition of a few steadily rotating waves.

The appearance of a disc having multiple spiral patterns, perhaps not
all having the same $m$-fold symmetry, would be more complicated.  But
there can be no doubt that the resulting superposed spiral arms would
also appear to wind up, since all that is required is that those
patterns closer to the disc centre should have higher pattern speed,
which is always true.  We emphasize this point in \S\ref{sec.shear}.

\begin{figure}
\includegraphics[width=.9\hsize,angle=0]{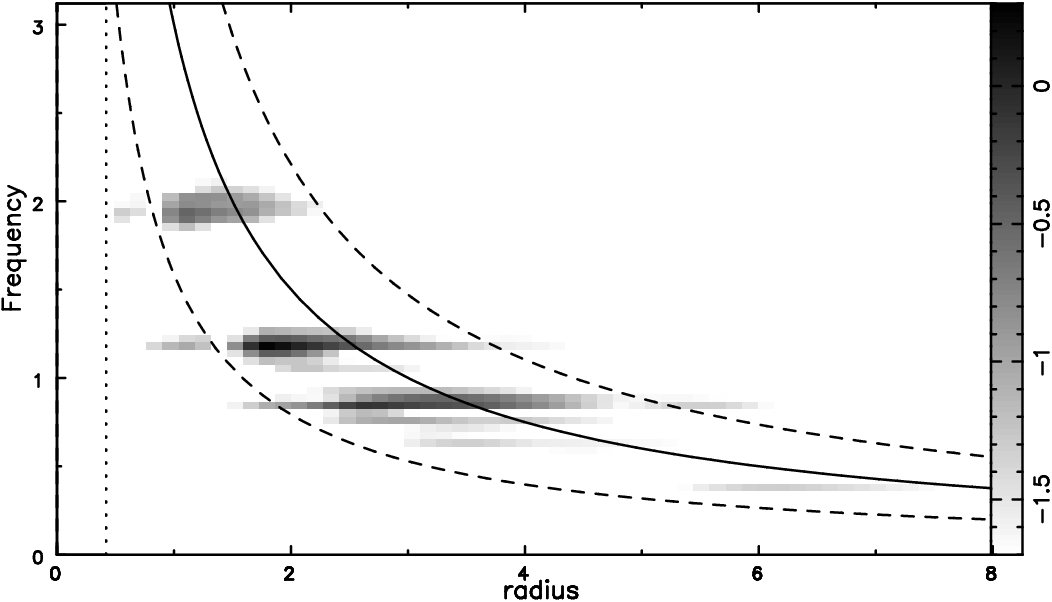}
% /home/sellwood/blobs/hi-spct2.s
\caption{As for fig.~\ref{fig.hi-spct} but for model 2H with the two
  heavier rings. This is computed over the time interval $50<t<200$.}
\label{fig.hi-spct2}
\end{figure}

\subsection{Heavier ring pair}
We next report an experiment with the masses of the forcing rings
increased ten-fold so that the forcing particles have the same masses
as in the heavier case with a single ring.  The amplitude evolution of
this coefficient from model 2H, green curve in
Figure~\ref{fig.two-rings} closely resembles that of the red line from
model 2L over the early part of the evolution, except that it is ten
times higher, but this ceases to be true after $t\sim60$.

\begin{figure}
\includegraphics[width=.9\hsize,angle=0]{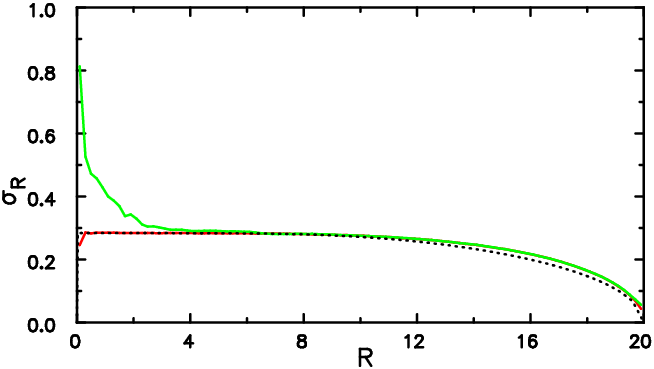}
% /home/sellwood/blobs/heat.s
\caption{The radial variation of the rms radial velocity at in the
  cases with two rings.  The black dotted line is at $t=0$ for both
  cases, while the coloured curves are at $t=200$; red for model 2L
  and green for model 2H.}
\label{fig.heat}
\end{figure}

The power spectrum from the time interval $50<t<200$ of model 2H is
presented in Figure~\ref{fig.hi-spct2}, where the driven responses to
the two forcing rings have frequencies $m\Omega_{p,1}=1.2$ and
$m\Omega_{p,2}=0.75$, but a third disturbance has appeared having
$m\Omega_{p,3}\la 2.0$.  This is almost certainly a groove mode that
has been excited by resonance scattering, since corotation for this
third pattern is close to the ILR of the higher forcing frequency.  It
is reasonable that this should be the first additional disturbance to
appear, since the dynamical clock runs faster at smaller radii.

The presence of this new instability causes the net amplitude of the
logarithmic spiral coefficient in model 2H, green curve in
Figure~\ref{fig.two-rings}, to depart from a scaled up version of
that from model 2L, red curve, at later times.  In fact, the
recognizably regular beats that characterize the red curve have become
something that looks almost chaotic when there are three dominant
disturbances.

The coloured lines in Figure~\ref{fig.heat} present the radial
velocity dispersion at $t=200$ in both models 2L (red line) and 2H
(green line).  Scattering at the ILR in model 2H has caused
significant heating in the inner disc; although the heated particles
have a narrow range of $L_z$ (\cf\ Fig.~\ref{fig.hi-diff}), they have
large $J_R$ and therefore significant radial velocities that dominate
in the very center where the unaffected disc is tapered away.
However, there was no change in $\sigma_R$ from the start (dotted
line) at any radius in model 2L, which is further evidence that
driving with the lower mass rings remained in the linear regime.

\section{Apparently shearing patterns}
\label{sec.shear}
Many authors (see \S\ref{sec.intro} for references) argue that spirals
are not density waves having fixed pattern speeds in the traditional
picture, but are instead shearing disturbances that wind up over time
at a rate close to that expected for material features.  We have
argued elsewhere, and continue to do so here, that the apparently
shearing spirals are merely the superposition of two or more rigidly
rotating density waves.

\begin{figure}
\includegraphics[width=.9\hsize,angle=0]{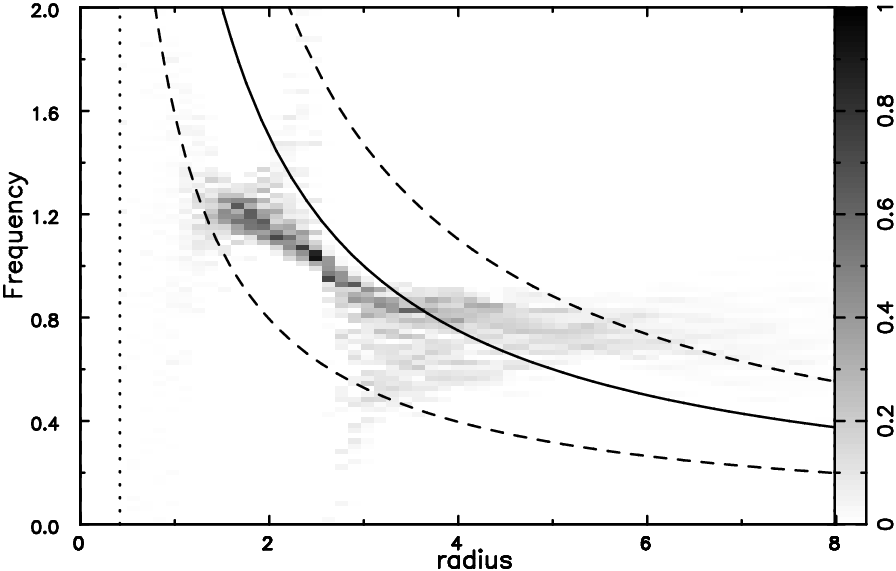}
% /home/sellwood/blobs/bswplt.s
\caption{The rate of change of the amplitude-weighted net phase of the
  two combined wake patterns discussed in \S\ref{sec.tworings}.  This
  figure is not a power spectrum of the kind shown in
  Figure~\ref{fig.hi-spct}, but results from the different procedure
  described in the text.  We have used data from the period $22<t<42$
  from model 2H, \ie\ a time interval before any additional
  disturbances appear.  The full drawn curve traces $m\Omega_c$ and
  the dashed curves $m\Omega_c \pm \kappa$.}
\label{fig.bswplt}
\end{figure}

In order to make this point more forcefully, we present
Figure~\ref{fig.bswplt}, which is drawn to resemble similar figures
presented by \citet{Baba13}, and we have followed their procedure to
create it.  They divide the disc into a set of rings and, at frequent
time intervals, they expand the relative disturbance density in
sectoral harmonics:
\begin{equation}
{\Sigma(R,\phi, t) \over \Sigma_0(R, t)} = \sum_{m=0}^\infty {\cal A}_m(R, t)
\cos\left[m\phi - \phi_m(R, t)\right],
\label{eq.sectanal}
\end{equation}
where ${\cal A}_m(R, t)$ is the amplitude and $\phi_m(R, t)$ the phase
of the $m$-th harmonic.  In our case, we have restricted disturbance
forces to the $m=3$ terms only, while they focused on $m=4$.  Following
their procedure, we compute a rough estimate of $m$ times the
``pattern speed'' from $[\phi_3(R, t_2) - \phi_3(R, t_1)] / (t_2-t_1)$
at two closely spaced times $t_1$ and $t_2$ at each radius throughout
a selected period of the evolution.  We then histogram the values
so obtained, weighting each by the amplitude ${\cal A}_3(R, t_2)$, and
present the results in Figure~\ref{fig.bswplt}.

The appearance of our Figure~\ref{fig.bswplt} resembles that in
figures 3 and 4 of \citet{Baba13}, in the sense that the dominant
values in this 2D histogram are strung out along a locus of decreasing
frequency with increasing radius that lies interior to the circular
orbit frequency, marked by the full-drawn curve.  The rough
first-order estimate of the pattern speed used in these plots
inevitably leads to some scatter in the plotted values, but outliers
are down weighted when the amplitude is low.  In our case the values
at smaller radii become roughly constant at the pattern speed of the
inner ring $m\Omega_p=1.2$, and there are hints of flattening again at
larger radii to values near $m\Omega_p=0.75$ for the outer ring.  But
the key point is that what we know to be the superposition of two
driven disturbances having distinct and radially constant frequencies,
appears in this plot, and in the movie shown in
Figure~\ref{fig.movie}, as a shearing pattern.

The relative amplitudes of the two waves (Fig.~\ref{fig.waves}) change
continuously over the radial range of the decline.  Since the faster
pattern dominates for $R\la2$ while the slower dominates at $R\ga3$,
the estimated frequency changes continuously in between.  We have
verified that doubling the radius, and therefore halving the
frequency, of the outer ring shifts the region of overlap of the two
waves to larger radii; in this case the declining ridge starts outside
corotation of the faster wave and drops more steeply to inside
corotation for the slower.  Thus the locations of ridges in figures
like \ref{fig.bswplt} is purely a consequence of the radial range over
which the relative amplitudes of the two disturbances changes.

Thus there can be no doubt that the superposition of two waves, can
appear as a shearing spiral.  In order to confirm the converse, that
an apparently shearing spiral results from the superposition of
separate patterns of fixed frequency, one must compute power spectra,
such as that shown in Figure~\ref{fig.hi-spct2}, which is computed
from a longer period in the same simulation (model 2H).  A power
spectrum is the temporal transform of the coefficients from the
surface density decomposition (eq.~\ref{eq.sectanal}); the longer the
time period used the higher the frequency resolution becomes, enabling
the combined density evolution to be separated into distinct peaks.
We have {\em always} found that this analysis decomposes the evolving
disturbance density variations into a small number of separate waves
having frequencies that are independent of radius.

\citet{Baba13} further claim nonlinear behaviour that caused some
particles in their simulations to gain random energy.  We agree that
wave-particle interactions that create random motion are
manifestations of nonlinear scattering that is expected to happen at
Lindblad resonances \citep{LBK}.  \citet{Baba13} argue that
because the spiral in their simulation is shearing, the heating must
be some new kind of nonlinear behaviour, but we suggest that the
shearing spiral they studied was in fact the superposition of two or
more density waves, and the heating they observed took place at the
Lindblad resonances of those steadily rotating patterns.

\citet{KN16} report experiments that were forced by co-orbiting rings
of particles in a similar manner to those here, although they
generally employed six particles per ring, as was more appropriate for
their very low mass discs.  The ``wakelets'' in their terminology were
the forced responses created swing amplification \citep{JT66, Bi20},
as we have also described above.  \citet{KN16} argued that the greater
amplitude when their wakelets from two rings align was a nonlinear
effect, but we have shown that the net amplitude fluctuates at the
beat frequency between the two disturbances, which peaks when the
ridge lines of the two separate waves align -- behaviour that can be
reconstructed by linear superposition (Figure~\ref{fig.movie}).  They
further report that the spirals created by the superposed wakelets are
shearing disturbances, but we have also shown that the apparent
evolution of the pitch angle of the combined response is reproduced by
the superposition of the two steady driven disturbances.

In summary, we are unconvinced by the alternative picture that these,
and many other, authors have embraced.  Here we have constructed
apparent shearing transients from the forced responses to two rings of
driving masses and shown (Figures~\ref{fig.movie} and
\ref{fig.bswplt}) that they have most of the properties that these
authors describe.  We have observed shearing transient spirals in most
of our simulations, and indeed were the first to report the phenomenon
\citep{SC84}, but power spectra of those same simulations \citep{Se89}
and every subsequent case have revealed that shearing transients
result from the superposition of a few steadily rotating waves.

Furthermore, we have reported \citep{SC14, SC19}, that the amplitudes
of spirals in our simulations do not change as the number of particles
is increased by several orders of magnitude, although it takes longer
to reach the common amplitude as $N$ is increased.  The most natural
explanation of this fact is that the spirals are linearly unstable
modes that exponentiate out of the noise, which is reduced as $N$ is
increased, and they saturate at a common amplitude due to nonlinear
effects.  The authors of papers who suggest that shearing transients
are the fundamental phenomenon point to swing amplification for their
origin but, because that mechanism merely amplifies an input signal by
a fixed amount, it would predict that the amplitudes of the resulting
spirals should decrease as $N^{-1/2}$, which is not what we have
reported.

Thus, no convincing mechanism has been presented in this alternative
view of the origin of spirals in simulations that can account for the
similar amplitudes of the spiral patterns as the number of disc
particles employed is varied by orders of magnitude.  Since we find
that power spectra always reveal a few discrete, steadily rotating
waves, we argue that spiral activity in simulations results from the
superposition of unstable modes \citep{SC14, SC19}.

\begin{figure}
\includegraphics[width=.9\hsize,angle=0]{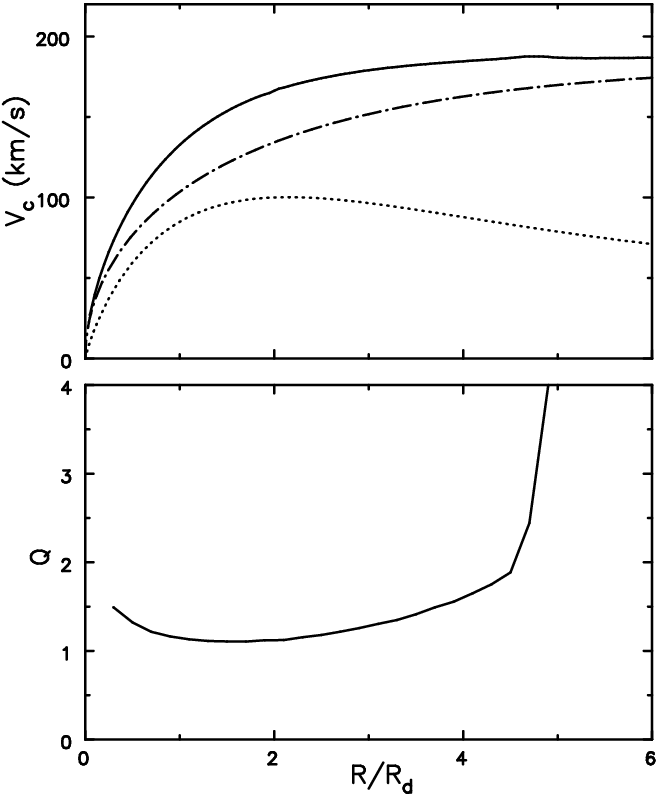}
% /home/sellwood/papers/SC20/figs/DVH0.s
\caption{Above: the rotation curve (full drawn), the disc (dotted),
  and halo (dot-dashed) contributions and below the $Q$ profile of our
  model intended to match that of \citet{DVH}}
\label{fig.DVH0}
\end{figure}

\section{A single perturbing mass}
\label{sec.DVH0}
In this section we present models having a single perturbing mass
and no imposed rotational symmetry.  

\subsection{A low mass disc}
We first match one of the experiments reported by \citet[][hereafter
  DVH]{DVH}, who employed an exponential disc having a scalelength
$R_d = 3.13\;$kpc and mass of $M_d = 1.9 \times 10^{10}\;$\Msun,
embedded in a Hernquist halo of mass $50M_d$ and radial parameter
$10R_d$.  The rotation curve is shown in the upper panel of
Figure~\ref{fig.DVH0}; the disc particle velocities were chosen to be
in rotational balance and to have the $Q$-profile shown in the lower
panel, with a minimum $Q\simeq 1.2$.  As usual, we prefer units in
which $G = M_d = R_d = 1$; our unit of velocity is therefore
$161.1\;$km~s$^{-1}$ and our unit of time is 19~Myr.

\begin{figure}
\includegraphics[width=.9\hsize,angle=0]{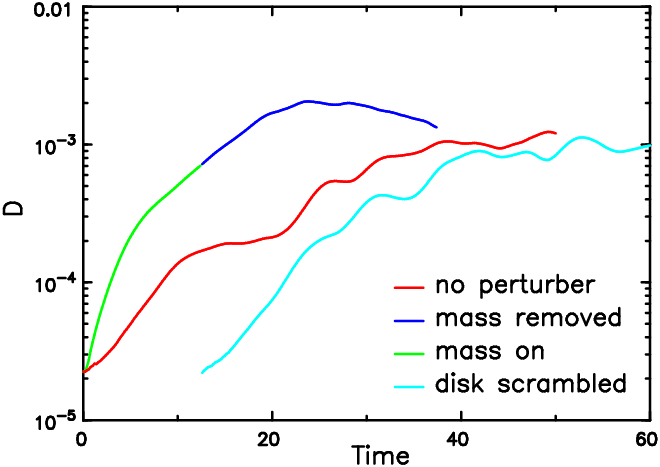}
% /home/sellwood/papers/SC20/DVHA.s
\caption{The time evolution of the rms relative mass variations on the
  entire grid in four separate simulations of models resembling those
  of \citet{DVH}.}
\label{fig.DVHA}
\end{figure}

\begin{figure}
\includegraphics[width=.9\hsize,angle=0]{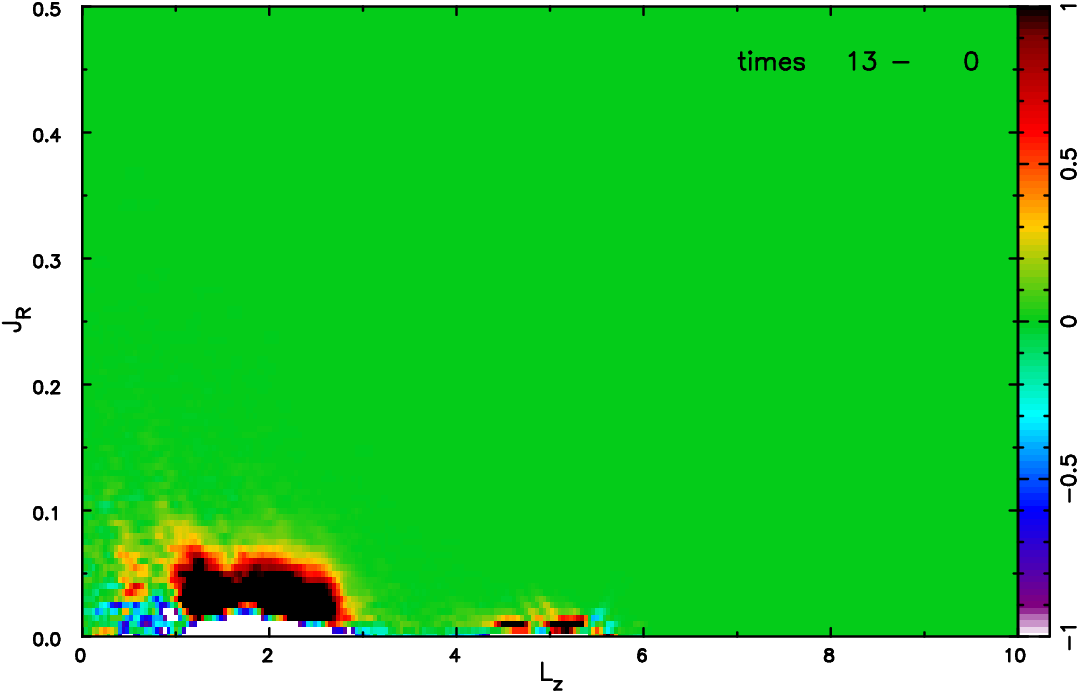}
% /home/sellwood/papers/SC20/figs/actdiff.s 4968 13 0
\caption{The phase space changes caused by a single perturbing mass in
  our reconstruction of the simulation having a single perturbing mass
  by \citet[][top left panel of their Fig.~10]{DVH}. The perturbing
  mass had $L_z = 2.06$ in our units.}
\label{fig.act4968}
\end{figure}

Since the disc of this model best supports smaller-scale structures
having rotational symmetries in the range $6 \la m \la 12$ (DVH),
we refined our 2D polar grid to $340 \times 512$, included sectoral
harmonics in the range $0 \leq m \leq 16$, adopted a shorter softening
length $\epsilon = 0.025R_d$ and doubled the number of particles to
match the number employed by DVH.

Following these authors, we introduced a single perturbing mass on a
circular orbit at $R=2R_d$ having a mass $M_d/2000$.  The orbit period
of this particle is $\sim 12.5$, and the wake that dressed the
perturber after one orbit closely resembled that shown in the upper
left panel of Fig.~10 of DVH.

For a quantitative measure of the total non-axisymmetric amplitude,
which is not confined to a single sectoral harmonic, we compute
\begin{equation}
D = \left\langle{ \left( \mu(i,j) - \bar \mu_j \right)^2} \right\rangle^{1/2} / M_d,
\end{equation}
where $\mu(i,j)$ is the mass of particles assigned to the grid point
$(i,j)$ and $\bar \mu_j$ is the average of the values on the $j\,$th
ring.  The quantity $D$ is therefore the rms relative mass variations
averaged over the entire grid. The green curve in
Figure~\ref{fig.DVHA} shows that non-axisymmetric mass variations
increase steadily in response to the perturbing mass.  The red curve
shows that non-axisymmetric variations also increase, but more slowly,
without a co-orbiting perturber, reflecting the onset of polarization
of the disc and possible mild instabilities, since this model is not
known to be stable.

The blue curve shows how $D$ evolves when the evolution is continued
after removing the perturber, when the disc manifested very similar
on-going activity of the kind reported by DVH.  However, the cyan line
reveals that activity grew scarcely more rapidly after scrambling this
disc than it did in the separate model without any perturbations (red
curve) and no clear frequencies stood out in the power spectra of its
evolution.

This null result implies that lasting changes wrought by the perturber
did not seed any vigorous new instabilities, and the continuing
activity shown by the blue line probably is caused by something like
the nonlinear mechanism proposed by DVH.  In order to understand why
the modified disc has no new instabilities, we present
Figure~\ref{fig.act4968}, which shows the phase space changes caused
by the single orbit of this perturbing mass.  Since Lindblad
resonances lie ever closer to corotation as $m$ increases, the
dominant higher sectoral harmonics caused the wake to have both a
smaller radial extent and the resonances to blur together.  Thus we
see that resonant scattering took place over almost the entire region
affected by the wake.  Since a groove of a fixed depth excites a
instability only when sufficiently narrow \citep{SK91}, the deficiency
of nearly circular orbits over a broad range of angular momentum
(Figure~\ref{fig.act4968}) is not destabilizing.

\subsection{Mestel disc again}
Responses to perturbations in the previous low-mass disc model are
most vigorous for sectoral harmonics $6 \la m \la 12$.  In
this circumstance, the strongest Lindblad resonance scattering from a
co-orbiting point mass takes place close to the radius of the
perturber and the resonances are sufficiently close together that the
separate scattering peaks from each resonance are blurred together, as
shown in Figure~\ref{fig.act4968}.  In order to create scattering
features that are not blurred together, we must employ a disc model
that responds more vigorously to lower sectoral harmonics, for which
the Linbdlad resonances will lie farther from the perturber and be
more widely spaced -- \ie\ a more massive disc.

We therefore returned to using the half-mass Mestel disc, in which we
inserted a single co-orbiting mass.  Since the swing-amplification
parameter $X=4/m$ in the half-mass Mestel disc, it is sufficient to
include sectoral harmonics $1 \leq m \leq 5$ only in order to capture
almost the entire self-consistent response \citep{To81, BT08}.  We
inserted a perturbing mass $M_p = 0.0025$ at $R_p = 5$, which has an
orbit period of $10\pi$.

\begin{figure}
\includegraphics[width=.9\hsize,angle=0]{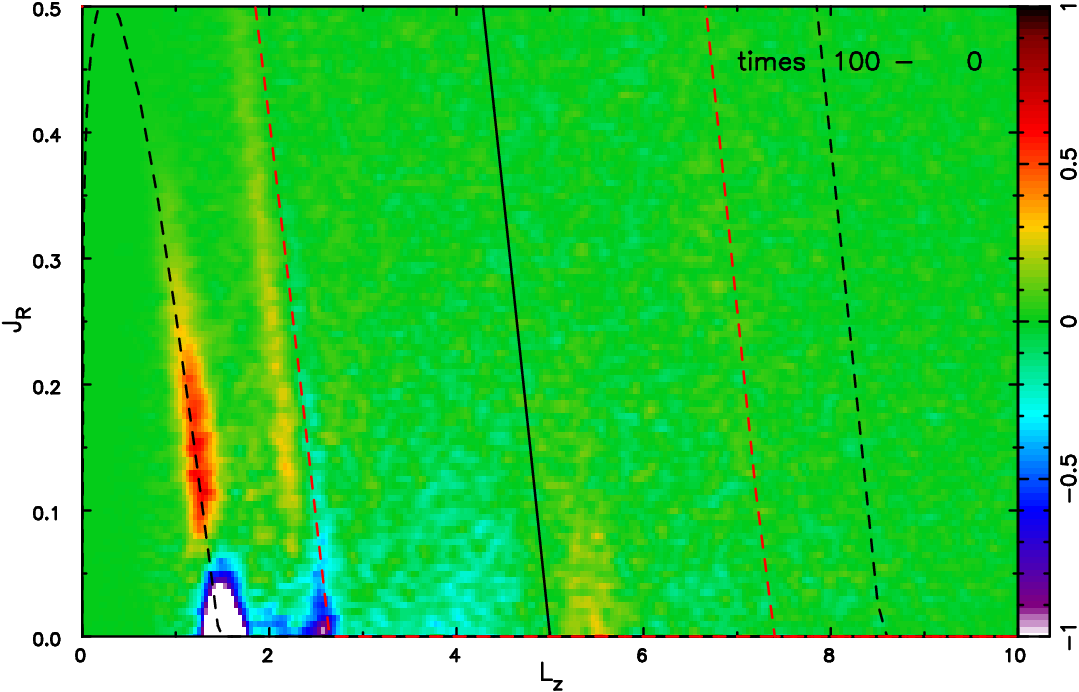}
% /home/sellwood/papers/SC20/figs/actdiff.s 4983 100 0
\caption{The phase space changes caused by a single perturbing mass at
  $R=5$ in the half-mass Mestel disc.  The solid line marks the locus
  of co-rotation, the dashed the Lindblad resonances for $m=2$ (black)
  and $m=3$ (red).}
\label{fig.act4983}
\end{figure}

The mild density response to this low-mass perturbation was barely
visible.  However, by $t=100$, or $\sim 3$ full orbits, the resulting
phase space changes, Figure~\ref{fig.act4983}, reveal strong
scattering at the $m=2$ ILR, weaker scattering at the $m=3$ ILR, some
changes near CR that result from trapping as the response to the
perturber has grown (see \S\ref{sec.1H} for a fuller explanation), but
little scattering at the OLRs.

\begin{table}
\caption{Dominant instabilities}
\label{tab.modes}
\begin{tabular}{@{}lccc}
 & $m$ & frequency & corotation \\
1 & 2 & $ 1.251\pm0.025 + (0.038\pm0.005)i$ & $1.60\pm0.03$ \\
2 & 2 & $ 0.716\pm0.003 + (0.015\pm0.010)i$ & $2.79\pm0.01$ \\
3 & 2 & $ 0.434\pm0.005 + (0.016\pm0.005)i$ & $4.61\pm0.05$ \\
4 & 3 & $ 1.110\pm0.003 + (0.025\pm0.005)i$ & $2.70\pm0.01$ \\
\end{tabular}
\end{table}

Continuing the evolution after removing the perturber, and scrambling
the azimuthal coordinates to disperse the wake, we found that the
model now possessed several unstable modes.  The dominant ones are
listed in Table~\ref{tab.modes}; three are bisymmetric and one has
three-fold symmetry.  The last column lists the implied corotation
radius, which in each case is close to a deficiency of near-circular
orbits visible in Figure~\ref{fig.act4983}.

This result establishes that scattering by isolated, co-orbiting
perturbing masses can seed groove instabilities in heavier discs.

\section{Conclusions}
\label{sec.concl}
In this paper, we have presented a study of spiral responses in
simulations that employed a simplified thin disc model of the stellar
component of a galaxy having an exactly flat rotation curve.  We
selected this model because a normal mode analysis of the smooth
stellar fluid by \citet{To81} had predicted it to be stable.  Although
the behaviour in simulations is much richer because of swing
amplification of the unavoidable shot noise from particles, we were
able effectively to suppress particle noise by employing large numbers
of particles.  We then studied the consequences of external forcing
that was both linear, for mild forcing, and nonlinear for slightly
heavier perturbing masses.  Here, nonlinearities are significant
changes to the phase space density of the particles caused by
resonance scattering.

At first, we forced our models with sets of three co-orbiting heavy
particles arranged symmetrically around rings and also restricted
self-gravity forces from the disc particles to the $m=3$ sectoral
harmonic.  The forcing masses quickly induced spiral responses in the
manner predicted by \citet[][see also Binney 2020]{JT66} that reached
nearly full amplitude after one full orbit, but continued to grow more
slowly thereafter.  When each perturbing mass was just $\sim 2000$
times more massive than the individual disc particles, we observed the
mild response to disperse with no lasting effects when the forcing
masses were removed after almost three orbits.  This behaviour is
possible only in a disc that is stable; forcing masses in a disc
having even mild instabilities would seed their growth, which would
continue after the forcing masses were removed.

Heavier perturbing masses, in both single rings and in two separate
rings, produced lasting changes to the disc that later caused
additional self-excited patterns to develop.  These new instabilities,
which were not present in the unperturbed disc, resulted from
nonlinear scattering of particles at the principal resonances of the
original driving perturbations and they persisted after we removed the
perturbations.

Our experiments with two rings of perturbing masses were particularly
illuminating, as they manifested the kind of shearing transient
spirals that have been reported in many recent papers, cited in the
introduction.  We were able to demonstrate in the movie
(Figure~\ref{fig.movie}) that the two separate steady responses shown
in Figure~\ref{fig.waves} could be superposed to create the evolving
disturbance density that appeared to wind up for a while and then
reform at the beat period.  The responses with heavier forcing masses
were proportionately stronger at first, and we analyzed this period by
the method employed by \citet{Baba13} (Figure~\ref{fig.bswplt}), which
revealed that the net disturbance density behaved as all these authors
argue even though it resulted from the superposition of two responses
driven at different frequencies.  Only power spectra
(\eg\ Figure~\ref{fig.hi-spct}) are able to resolve the evolving
disturbance density into separate coherent, rigidly rotating patterns
-- behaviour that we have always found in our work.

We also exposed the connection between our work and that of
\citet{DVH}.  We reproduced their simulation with a single co-orbiting
mass, in which the response in their low mass disc was dominated by
sectoral harmonics in the range $6 \la m \la 12$.  Since Lindblad
resonances lie ever closer to corotation as $m$ increases, the wake
both had a small radial extent and resonant scattering took place over
almost the entire region affected by the wake, creating a deficiency
of circular orbits over broad range of $L_z$.  We found no new
instabilities were present after we removed the perturber, consistent
with the prediction \citep{SK91} that broader grooves are not expected
to be destabilizing.  Galaxy discs generally prefer 2- and 3-arm
spirals \citep{Davi12, Hart16, Yu18}, which is indicative of heavier
discs.  We therefore also presented experiments with the half-mass
Mestel disc in which swing amplification favors $m=2$ and 3, and
showed that a single perturbing mass caused narrow scattering features
at the ILRs of these sectoral harmonics that did excite new linear
instabilities.

In this, and previous papers \citep{SC14, SC19}, we have presented
compelling evidence that spirals develop as a result of linearly
unstable modes in a non-smooth disc, and that nonlinear scattering at
resonances is responsible for subsequent instabilities.  The unstable
modes are disturbances having a fixed frequency over a broad range of
radii from the inner to (generally) the outer Lindblad resonance.
Here, we have shown that the popular alternative shearing spiral
description is no more than the consequence of the superposition of
two or more steadily rotating patterns.  We have confirmed that
something like the mechanism proposed by \citet{DVH} does in fact
operate in unrealistically low-mass discs only, but new instabilities
are again provoked by resonance scattering caused by a single
co-orbiting mass in realistically heavy discs.  Thus, we believe we
now fully understand the origin of spiral patterns in simulations of
isolated disc galaxies.  Whether it is also the origin of spirals in
nature will be harder to establish.

\section*{Acknowledgements}
We thank Luis Ho and an anonymous referee for comments on the
manuscript.  JAS acknowledges the continuing hospitality of Steward
Observatory.

\section*{Data availability}
The data from the simulations reported here can be made available
on request.


% Don't change these lines
\bsp	% typesetting comment
\label{lastpage}
\end{document}